\documentclass[preprint,aps,showpacs]{revtex4}
\usepackage{graphicx}
\usepackage{amsmath}
\usepackage{amssymb}
\usepackage{hyperref}
\usepackage{latexsym}
\usepackage{epsfig}
\usepackage{array}

 

\begin{document}

\title{Dipolar collisions of polar molecules in the quantum regime}

\author{K.-K. Ni,$^{1\ast}$ S. Ospelkaus,$^{1\ast}$ D. Wang,$^{1}$ G. Qu\'em\'ener,$^{1}$ B. Neyenhuis,$^{1}$ \\M. H. G. de Miranda,$^{1}$ J. L. Bohn,$^{1}$ J. Ye,$^{1\dagger}$ D. S.
 Jin$^{1\dagger}$\\}
\affiliation{
\normalsize{$^{1}$JILA, NIST and University of Colorado,}\\
\normalsize{Department of Physics, University of Colorado}\\
\normalsize{Boulder, CO 80309-0440, USA}\\
\normalsize{$^\ast$These authors contributed equally to this work; }\\
\normalsize{$^\dagger$To whom correspondence should be addressed; }\\
\normalsize{E-mail:  Ye@jila.colorado.edu; Jin@jilau1.colorado.edu}
}

\maketitle

\textbf{ Ultracold polar molecules offer the possibility of
exploring quantum gases with interparticle interactions that are
strong, long-range, and spatially anisotropic.  This is in stark
contrast to the much studied dilute gases of ultracold atoms, which
have isotropic and extremely short-range, or  ``contact",
interactions. Furthermore, the large electric dipole moment of polar
molecules can be tuned with an external electric field;  this
provides unique opportunities such as the control of ultracold
chemical reactions~\cite{Krems}, a unique platform for quantum
information processing~\cite{DeMille2002,DeMille,Yelin}, and the
realization of novel quantum many-body
systems~\cite{Micheli,Lahaye,Baranov,Pupillo}. In spite of intense
experimental efforts aimed at observing the influence of dipoles on
ultracold molecules~\cite{reviewpolar}, only recently have
sufficiently high densities been achieved \cite{polarmol}. Here, we
report the experimental observation of dipolar collisions in an
ultracold molecular gas prepared close to quantum degeneracy. For
modest values of an applied electric field, we observe a dramatic
increase in the loss rate of fermionic KRb molecules due to ultrcold
chemical reactions. We find that the loss rate has a steep power-law
dependence on the induced electric dipole moment, and we show that
this dependence can be understood with a relatively simple model
based on quantum threshold laws for scattering of fermionic polar
molecules. In addition, we directly observe the spatial anisotropy
of the dipolar interaction as manifested in measurements of the
thermodynamics of the dipolar gas. These results demonstrate how the
long-range dipolar interaction can be used for electric-field
control of chemical reaction rates in an ultracold polar molecule
gas. Furthermore, the large loss rates in an applied electric field
suggest that creating a long-lived ensemble of ultracold polar
molecules may require confinement in a two-dimensional trap geometry
in order to suppress the influence of the attractive head-to-tail
dipolar interactions~\cite{Micheli2D,Buchler,Li,Quemener}.}

Dipolar interactions have been explored in several atom gas
experiments using the magnetic dipole moments of
atoms~\cite{Pfau,Stamper-Kurn}, however this interaction is
intrinsically orders of magnitude weaker than the dipolar
interaction between typical polar molecules. Ultracold gases of
polar molecules, then, open the possibility for realizing strong,
and therefore relatively long-range, interactions. For example, with
polar molecules confined in optical lattice potentials, one could
realize a system where the interactions between particles in
neighboring sites is as strong as the on-site interactions now
commonly realized with atoms. This longer range interaction for
polar molecules will allow access to a new regime of strongly
correlated quantum gases with unique phase transitions, such as to
supersolid phases for bosons \cite{Goral,Capo} and to topological
superfluid phases for fermions~\cite{pairing}. Another important
difference between magnetic and electric dipolar interactions is
that the strength of the effective electric dipole moment is tunable
with an applied electric field. In addition to its obvious utility
for controlling the interaction strength in dipolar quantum gases,
the electric-field dependence could be exploited in the development
of novel quantum computing schemes or in the control of ultracold
chemical reactions.

\begin{figure}
   \includegraphics[width=14 cm]{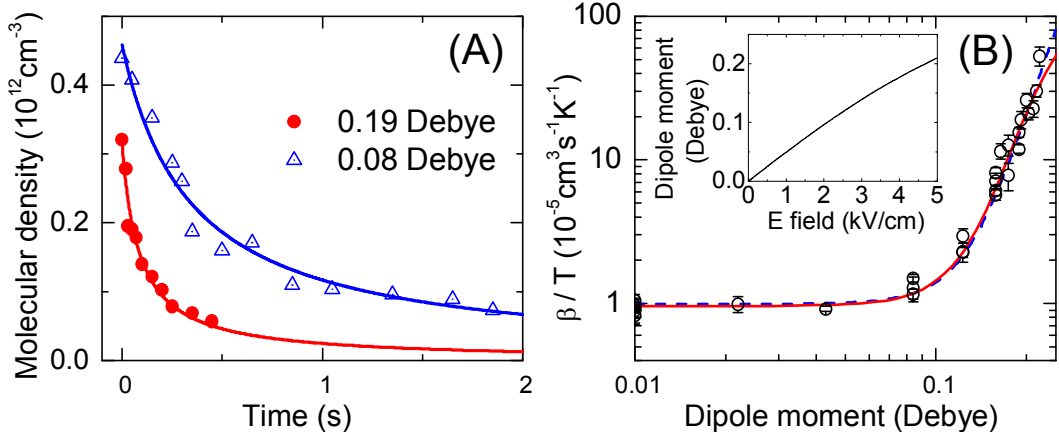}
 \caption{Two-body inelastic loss for fermionic polar molecules.  (A) We extract the inelastic loss rate coefficient
 $\beta$ from a fit (solid lines) to the measured time evolution of the trapped molecular gas density.
 The data shown here are for an induced dipole moment $d= 0.08$ Debye (open triangles) and
 $d=0.19$ Debye (filled circles), and an initial temperature $T=300$ nK.  (B) We plot
 $\beta$, divided by $T$, as a function of the
 induced dipole moment, $d$. A fit to a power-law dependence in the region $d>0.1$ Debye yields
 a power of $6.1\pm0.8$. The dashed line shows a fit to a simple model based on the quantum
 threshold behavior for tunneling through a dipolar-interaction modified $p$-wave barrier (see text).
 The solid line shows the result of a more complete quantum
 scattering
calculation including an absorbing potential at short range. Note
that the full model (dashed line) deviates from the simple model
prediction at our largest $d$ where the scattering is no longer in
the Wigner threshold regime. The calculated dependence of the
induced dipole
 moment on the applied electric field is shown in the inset to (B).} \label{fig1}
\end{figure}

We perform our experiments with an ultracold gas of
$^{40}$K$^{87}$Rb molecules prepared in a single nuclear hyperfine
state within the ro-vibronic ground state
($^1\Sigma^+$)~\cite{polarmol, hyperfine}.  The gas is confined in a
pancake-shaped optical dipole trap, which is realized by overlapping
two horizontally propagating, elliptically shaped laser beams with a
wavelength of $\lambda=1064\, \mathrm{nm}$. Typical harmonic
trapping frequencies are $\omega_\textrm{x}=2\pi\times 40\;
\textrm{Hz} $ and $\omega_\textrm{z}=2\pi\times 280\; \textrm{Hz}$,
in the horizontal and vertical directions, respectively.
The KRb molecules have a permanent electric dipole moment of 0.57
Debye~\cite{polarmol}, where 1 Debye = $3.336 \times 10^{-30}$ C m.
However, the effective molecular dipole moment in the lab frame is
zero in the absence of an external electric field. When an external
electric field is applied, the molecules begin to align with the
field, and have an induced dipole moment, $d$, that increases as
shown in the inset of Fig.~\ref{fig1}~B. This figure covers the
range of applied electric field that we currently access
experimentally, which corresponds to an accessible dipole moment
range of $0$ to $0.22\,\textrm{Debye}$. In our setup, the external
electric field points in the vertical direction ($\hat{z}$),
parallel to the tight axis of the optical trap. Thus, the spatially
anisotropic dipolar interactions will be predominantly repulsive for
molecules colliding in the horizontal direction (side-by-side) and
predominantly attractive for molecules approaching each other along
the vertical direction (head-to-tail).

In an ultracold gas, the quantum statistics of the particles plays
an essential role in the interactions.  Our $^{40}$K$^{87}$Rb
molecules are fermions prepared in a single internal quantum state
at a temperature equal to 1.4 times the Fermi temperature.
Therefore, the quantum statistics requires that the wave function
for two colliding molecules be antisymmetric with respect to
molecule exchange.  If one considers the relative angular momentum
between two colliding molecules, this means that scattering can only
proceed via odd partial waves, and will be dominated by angular
momentum $L=1$ ($p$-wave) scattering at ultralow temperature.
Previous work at zero electric field (without long-range dipolar
interactions) showed that the lifetime of the trapped
$^{40}$K$^{87}$Rb molecules is limited by atom-exchange chemical
reactions that proceed via $p$-wave scattering
~\cite{ultracoldchem1}. In our experiments, the typical
translational temperature of the molecular gas is 300 nK, while the
energy height of the $p$-wave barrier for $^{40}$K$^{87}$Rb
molecules corresponds to a temperature of approximately 24
$\mu$K~\cite{Kotochigova}.  With the barrier height much larger than
typical collision energies, scattering rates in the molecular cloud
are determined by the tunneling rate through the centrifugal barrier
and the molecular gas lifetime is relatively long (on the order of 1
s)~\cite{ultracoldchem1}.

In this paper, we investigate the effect of electric dipolar
interactions on collisions and find a surprisingly large effect even
for our relatively modest range of applied electric fields.  We
measure the molecular loss rate by monitoring the time evolution of
the average number density of trapped molecules, $n$ (see
Fig.~\ref{fig1}~A). We fit the data to the solution of
 \begin{eqnarray}
\frac{d n}{d t}=-\beta n^2 - \alpha n,
\end{eqnarray}
shown as solid lines in Fig.~\ref{fig1}~A.

The first term on the right hand side accounts for number loss, and
we extract the measured two-molecule inelastic loss rate
coefficient, $\beta$, (which is two times the collisional event
rate) from the fit. The second term describes density change arising
from heating of the trapped gas during the measurement.  In a single
measurement, we observe an increase in temperature that is at most
50$\%$.  In subsequent analysis, we fit the measured temperature to
a linear heating and obtain a constant slope $c$.  In Eqn. 1, we
then use $\alpha=\frac{3}{2}\frac{c}{T+c t}$, where $T$ is the
initial temperature of the gas (see also \cite{ultracoldchem1}).

Fig.~\ref{fig1}~B shows a summary of our experimental data, where we
plot $\beta/T$, as a function of $d$. We plot the ratio $\beta/T$
because the Wigner threshold law for $p$-wave scattering predicts
that $\beta$ scales linearly with $T$, and we previously verified
this temperature dependence at $d=0$ Debye \cite{ultracoldchem1}.
For the data in Fig.~\ref{fig1}, $T$ ranged from 250 nK to 500 nK.
In Fig.~\ref{fig1}~B we see that dipolar interactions have a
dramatic effect on the inelastic collision rate. At low electric
field, where $d$ is below 0.1 Debye, we observe no significant
modification to the zero electric-field loss rate (which is plotted
at $d=0.01$ Debye for inclusion on the logarithmic scale). However,
for higher electric fields, we observe a rapidly increasing loss
rate, with well over an order of magnitude increase in $\beta/T$ by
$d=0.2$ Debye.  Fitting the data for $d>0.1$ Debye, we find that the
inelastic rate coefficient follows a power-law dependence on $d$,
$\beta/T\propto d^p$, where $p=6.1\pm0.8$.

\begin{figure*}
   \includegraphics[width=14 cm]{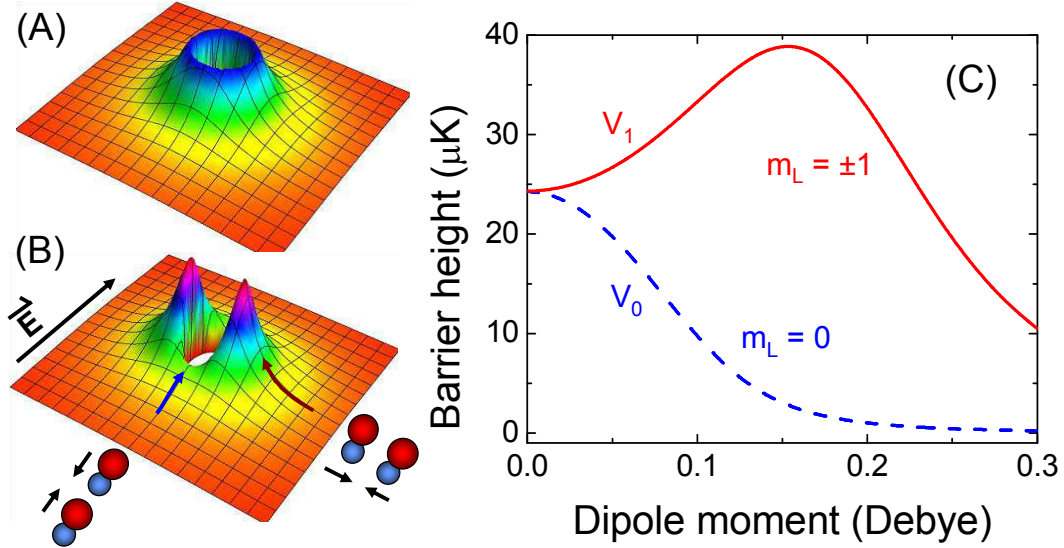}
 \caption{$p$-wave centrifugal barrier for dipolar collisions between fermionic polar molecules.
 (A) Schematic picture of the effective intermolecular potential for fermionic molecules at zero electric
 field.  At intermediate intermolecular separation (center of plot), two
 colliding molecules are repelled by a large centrifugal barrier for $p$-wave collisions.
(B) Schematic picture of the effective intermolecular potential for
a relatively small applied electric field.  The dipolar interactions
are spatially anisotropic, with an attraction that reduces the
barrier for head-to-tail collisions and a repulsion that increases
the barrier for side-by-side collisions. (C) Height of the $p$-wave
barrier as a function of dipole moment for repulsive dipole-dipole
interactions ($V_1$, solid red line) and for attractive
dipole-dipole interactions ($V_0$, dashed blue line).  As the
induced dipole moment increases, the spatially anisotropic dipolar
interactions lower the centrifugal barrier for $m_L=0$ collisions
and raise the barrier for $m_L=\pm1$ collisions. The lowering of the
$m_L=\pm 1$ barrier at very large dipole moments is due to mixing
with higher partial waves, e.g. $L=3,5,7,....$.} \label{fig2}
\end{figure*}

To understand this strong electric-field dependence of the inelastic
loss rate, we consider a relatively simple quantum tunneling model
where the loss is assumed to be due to collisions between fermionic
molecules that proceed via tunneling through a $p$-wave centrifugal
barrier followed by unit probability for loss at
short-range~\cite{Quemener}. The fact that we do not observe any
resonant oscillations as a function of E-field (see Fig. 1 B) is
consistent with a very high loss probability for molecules reaching
short range. In an applied electric field, the long-range
dipole-dipole interaction $\propto \frac{1}{R^3}$, where $R$ is the
intermolecular separation, significantly modifies the height of the
$p$-wave barrier, and thus changes the inelastic collision rate.
Moreover, the fact that the dipole-dipole interaction is spatially
anisotropic means that the $p$-wave barrier height will be different
for $m_L=0$ and $m_L=\pm1$ scattering, where $m_L$ is the projection
of the relative orbital angular momentum $L$ onto the electric-field
direction.  In particular, the attractive nature of dipole-dipole
interaction for polar molecules colliding ``head-to-tail'' lowers
the barrier for $m_L=0$ collisions, while the repulsive
dipole-dipole interaction for polar molecules colliding
``side-by-side'' raises the barrier for $m_L=\pm1$ collisions.
Fig.~\ref{fig2}~A,B show these effects schematically, while
Fig.~\ref{fig2}~C shows the calculated maximum height of the $m_L=0$
and $m_L=\pm1$ collisional barriers, $V_0$ and $V_1$, respectively,
as a function of $d$.

In our simple model, we assume that the collision rate follows the
Wigner threshold law for $p$-wave inelastic collisions, namely
$\beta \propto T/V_{0,1}^{3/2}$.  For large $d$, $V_0$ is
significantly smaller than $V_1$, and the loss will proceed
predominantly through ``head-to-tail" attractive collisions of the
polar molecules. In this regime, $V_0$ scales as $d^{-4}$ and the
model predicts that $\beta/T$ will increase with a characteristic
power law of $d^6$ for $d>0.1$ Debye \cite{Quemener}.  This
prediction is in excellent agreement with our measured dependence of
the loss rate on $d$ for $d>0.1$ Debye (Fig. 1 B).

For a quantitative description of the inelastic collisional rate
over our full range of experimentally accessible dipole moments, we
include contributions from both $m_L=0$ and $m_L=\pm 1$ collisions,
and we calculate the barrier heights using adiabatic potential
curves that include mixing with higher partial waves (see Fig. 2 C).
We fit the prediction of this quantum threshold model to our data
using a scaling factor, $\gamma$, as a free fit parameter, where
$\gamma$ can be interpreted as the loss probability when the
collision energy equals the height of the barrier. The resulting
theoretical prediction (solid line in Fig.~\ref{fig1}) agrees very
well with our experimental data (open circles), and we obtain
$\gamma=0.35\pm0.08$. However, in order to get this agreement over
the full range of $d$, we found it necessary to introduce a second
fit parameter that multiplies the coefficient, $C_6$, of the
van-der-Waals interaction. The best fit value of this parameter is
$1.9\pm0.9$. Also shown in Fig.~\ref{fig1}~B as a dashed line is the
result of a more complete quantum scattering calculation that
employs a strong absorptive potential at short range but captures
the long-range physics and uses $C_6$ as the single fit parameter.
This fit also agrees well with the experimental data, and gives
$C_6=21000\pm7000$ a.u. in good agreement with the predicted value
of $C_6=16130$\,a.u.~\cite{Kotochigova}.

\begin{figure}
   \includegraphics[width=10 cm]{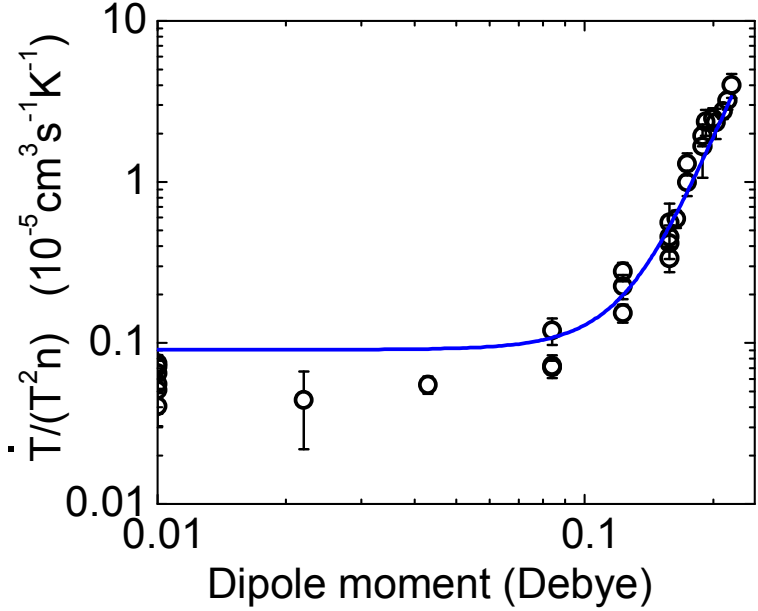}
 \caption{Normalized fractional heating rate $\frac{\dot{T}}{T}/(Tn)$ as a
 function of dipole moment. The heating rate is extracted using a linear fit
 to the initial temperature increase and is then normalized
  by the initial density and temperature of the ensemble. The line is the expected
  heating rate given by $\dot{T}/(T^2n)=(\beta/T) /12$ (see text).  Typical initial conditions
for these data are $n=0.3\cdot 10^{12}/~\textrm{cm}^3$ and
$T=0.5~\mu \textrm{K}$, and the absolute heating rate ranges from
$0.1~\mu\textrm{K}/\textrm{s}$ at zero electric field to
$2~\mu\textrm{K}/\textrm{s}$ at our largest electric fields. }
\label{fig3}
\end{figure}

Accompanying the increased inelastic loss rates for increasing $d$,
we observe an increased heating rate for the polar molecule gas. In
Fig.~\ref{fig3} we plot the measured initial heating rate as a
function of $d$. The heating rate $\dot{T}=c$ is extracted using a
linear fit to the measured molecular cloud temperature vs. time,
over a time period sufficiently long to allow $T$ to increase by
approximately $20$ to $30\%$. In Fig~\ref{fig3}, we plot the
fractional heating rate $\dot{T}/T$ normalized by both the initial
$n$ and the initial $T$. We have developed a simple thermodynamic
model for heating that is directly caused by the inelastic loss. We
consider the energy lost from the gas when molecules are removed in
inelastic collisions, and assume that the gas stays in thermal
equilibrium. In this model (see Supplementary Information), the
heating arises solely from an``anti-evaporation'' mechanism in the
trap~\cite{Roberts,Weber}, where the particles removed by inelastic
collisions have, on average, lower energies than typical particles
in the gas. One way to understand this heating mechanism is simply
to note that inelastic collisions preferentially remove particles
from the center of the trap where the number density is the highest,
and where the particles have the lowest potential energy from the
trap. We also include in our calculations a competing
``cooling''effect that comes from the fact that the $p$-wave
inelastic collision rate increases linearly with the collision
energy. Including these two competing effects, we obtain
$\dot{T}/(T^2 n)=(\beta/T)/12$ (see Supplementary Information).
Remarkably, this simple prediction, Fig.~\ref{fig3} (solid line),
using the $\beta/T$ values from the fit to our loss rate data in
Fig.~\ref{fig1} with no additional free parameter, agrees very well
with the independently measured heating rates for the molecule gas.

The anisotropy of the dipole-dipole interaction is directly revealed
in an anisotropic distribution of molecules in the trap. The average
energy per particle, which we measure from the expansion of the gas
following a sudden release from the trap, can be different in the
vertical and horizontal directions. In the following, we present
measurements of the time evolution of the expansion energy in the
two directions for different $d$. To probe the spatial anisotropy of
dipolar collisions, we start by adding energy along one direction of
the cylindrically symmetric trap using parametric heating. Here, we
modulate the power of both optical trapping beams at twice the
relevant harmonic trapping frequency for 50~ms (z-direction) or
100~ms (x,y-directions). We then wait 100~ms before quickly
increasing the electric field (in less than $1\mu$s) to the desired
final value and measuring the time dependence of the vertical
($T_z$) and horizontal ($T_{x}$) ``temperatures'' of the cloud.
Here, $T_z$ and $T_{x}$ simply correspond to the measured expansion
energies in the two directions. We note that this type of
measurement is commonly used in ultracold atom gas experiments to
measure the elastic collision cross section~\cite{retherm}.

\begin{figure*}
   \includegraphics[width=14cm]{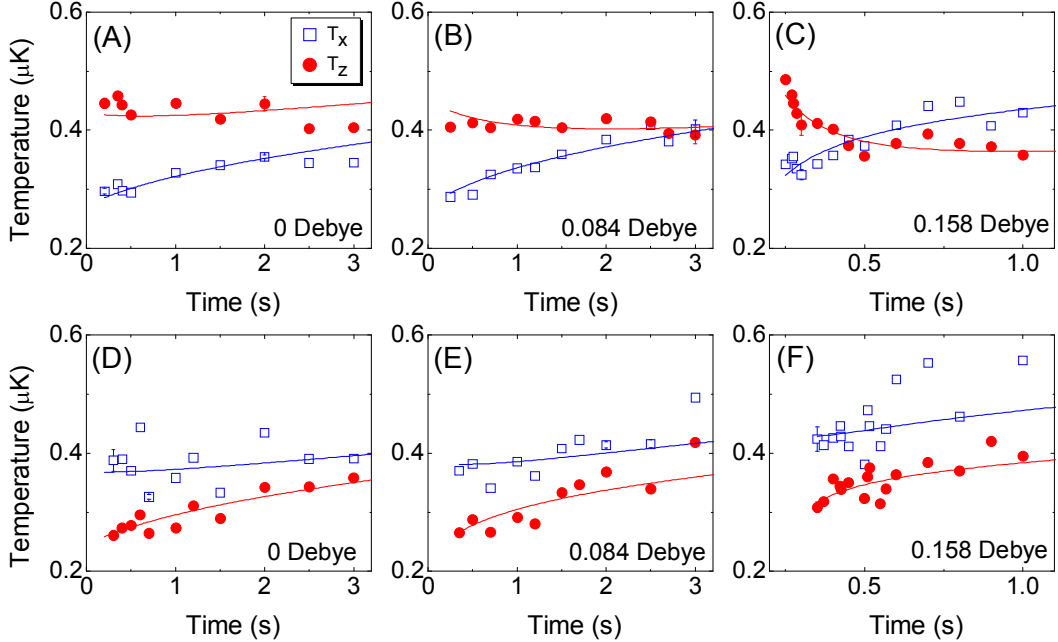}
 \caption{ Apparent cross-dimensional rethermalization as a function of dipole
  moment for $T_z>T_x$ (upper row, A-C) and $T_z<T_x$ (lower row, D-F) in the polar molecule gas.
The experimental data reveal a striking difference between the
results for heating the gas in the vertical direction (A to C) and
heating in the horizontal directions (D to F), and thus provide
evidence for the strong anisotropic characteristic of dipolar
interactions (see text). The electric field is applied along
$\hat{z}$.} \label{fig4}
\end{figure*}

Figure~\ref{fig4} shows the experimental data for these
rethermalization experiments with three values of $d$ and under two
initial conditions: $T_z>T_{x}$ (upper row, A-C) and $T_{z}<T_x$
(lower row, D-F). For $d=0$ Debye (Fig.~\ref{fig4} A and D), $T_z$
and $T_{x}$ equilibrate very slowly on a timescale of approximately
$4$~s. Because $d=0$ Debye, there are no dipolar interactions and
the data agree with our expectation of very slow equilibration for
spin-polarized fermions. Indeed, this is the longest
rethermalization time we have observed in our trap, and therefore
the data are consistent with no elastic collisions and only
technical imperfections such as a small cross-dimensional coupling
in the trapping potential.

In an applied electric field, the elastic collision cross section
due to long-range dipolar interactions is predicted to increase as
$d^4$ \cite{Bohn}. For the case where initially $T_z>T_{x}$ (Fig.
4~B,C), our data show that $T_z$ and $T_x$ approach each other, in
what appears at a casual glance to be cross-dimensional
rethermalization. The timescale for this apparent rethermalization
even decreases steeply with increasing $d$ as one might expect.
However, we note that in Fig. 4~C the temperatures cross each other
at long times, which is inconsistent with rethermalization driven by
elastic collisions. Even more striking is the fact that the
thermodynamic behavior of the gas is completely different when the
gas initially has $T_z<T_x$.  In this case, $T_z$ and $T_x$ do not
equilibrate during the measurement time (see Fig.~\ref{fig4}~E,F).

 The explanation for these
surprising observations comes from the spatially anisotropic nature
of inelastic dipole-dipole collisions and the fact that the molecule
gas experiences number loss.  We have seen (Fig.~3) that loss due to
inelastic collisions heats the gas, and we can quantitatively
understand this heating rate by considering the effect of molecule
loss on the average energy per particle. We can adapt the heating
and inelastic collision model described above to allow the average
energy per particle, or ``temperature'', to be different in the two
trap directions. The ``anti-evaporation" picture then predicts that
the dominant head-to-tail collisions ($m_L=0$) will lead to heating
in $x$ and $y$ but cooling along $z$. Side-by-side collisions
($m_L=\pm 1$), on the other hand, should contribute to heating along
$z$ but give no temperature change in $x$ and $y$ (see Methods and
Supplementary Information). To compare this model to our
rethermalization-type data, we fix the $d$-dependent $\beta$ using
the fit to our data in Fig.~\ref{fig1} (solid line). This fixes both
the time evolution of the molecule number as well as the heating
rates in the two trap directions. We then accommodate possible
elastic collision effects in the model by adding a term that would
exponentially drive the energy difference between the two directions
to zero. Figure~\ref{fig4} shows a comparison between the results of
the model (solid lines) and our experimental data. Although the
model uses few free parameters (only the elastic collision
cross-section, in addition to the initial $n$, $T_x$, and $T_z$), it
provides excellent agreement with the experimental data.

Our observations provide clear evidence of the anisotropic nature of
the dipole-dipole interactions through the observed anisotropy in
the apparent rethermalization. From the agreement of the data with
the model, we conclude that the rethermalization behavior is
actually dominated by the anisotropic nature of the inelastic
collisions. What then can we conclude about elastic collisions? The
best fit value for the elastic collision cross section is finite and
increases with increasing $d$. This is consistent with the
prediction that the elastic collision rate for polar molecules will
scale as $d^4$ \cite{Bohn}, and, furthermore, the fit values agree
with the prediction of cross sections on the order of
$\sigma_{\textrm{el}}=7\cdot10^{-8}\textrm{cm}^2/(\textrm{s}~\textrm{Debye}^4)$~\cite{Bohn}.
However, the presence of inelastic loss and the resulting
anisotropic heating make it difficult to accurately extract a
measured value of the elastic cross section for a more precise
comparison with theory.

The results presented here demonstrate that modest applied electric
fields can dramatically alter the interactions of fermionic polar
molecules in the quantum regime. For future efforts aimed at
studying many-body phenomena in dipolar molecular quantum systems it
will be necessary to protect the gas from strong inelastic loss and
heating~\cite{Micheli2D,Buchler,Gorshkov}.  In particular, the
demonstration here of strong spatial anisotropy for inelastic
collisions of polar molecules suggests that going to a
two-dimensional trapped gas will be a promising route to realizing a
long-lived quantum gas of polar molecules with dipole-dipole
interactions. This could be achieved, for example, with polar
molecules confined in an array of 2D pancakes in a 1D optical
lattice trap \cite{Micheli2D,Buchler}.  Interestingly, even when
short-range inelastic loss processes are suppressed, the attractive
part of the long-range dipole-dipole interaction could still give
rise to correlations between neighboring 2D pancakes in the 1D
lattice trap~\cite{Demler,Santos}.

\subsection{Methods}

For the fit to a quantum threshold model~\cite{Quemener} in Fig.
1~B, we write the inelastic loss rate coefficient, $\beta =K_0 T_z +
2 K_1 T_x$, as the sum of two terms corresponding to $m_L=0$ and
$m_L=\pm1$ scattering, respectively, and we assume $T_z=T_x=T$. The
$d$-dependent coefficients, $K_0$ and $K_1$ are obtained using
\begin{eqnarray}
K_{0,1} &=& \gamma\frac{3 \, \pi \, \hbar^2 }{\sqrt{2 \mu^3}
V_{0,1}^{3/2}} \ k_B, \label{ratenum}
\end{eqnarray}
where $\hbar=\frac{h}{2\pi}$, $h$ is Planck's constant, and $k_B$ is
the Boltzmann constant. The barrier heights, $V_0$ and $V_1$, are
taken to be the maximum energy of the long-range adiabatic potential
$V(R)$ evaluated in a basis set of partial waves $|L M_L \rangle
$~\cite{Quemener}. The potential $V(R)$ includes a repulsive
centrifugal term, $\hbar^2 L(L+1) / (2 \mu R^2)$ where $\mu$ is the
reduced mass of the colliding molecules, an attractive isotropic van
der Waals interaction $-b C_6 / R^6$, and the dipolar interaction.
We use only two fit parameters, $b$ and $\gamma$, when fitting this
model to the measured $\beta/T$ vs. $d$.

The solid lines in Fig. 4 are a fit of the measured time evolution
of $n$, $T_z$, and $T_x$ to the numerical solution of three
differential equations (see Supplementary Information):
\begin{eqnarray}
\frac{dn}{dt}=-(K_0 T_z+2 K_1
T_x)n^2-\frac{n}{2T_z}\frac{dT_z}{dt}-\frac{n}{T_x}\frac{dT_x}{dt}\\
\frac{dT_z}{dt}=\frac{n}{4}(-K_0 T_z+2 K_1 T_x)T_z-\frac{2
\Gamma_{el}}{3}(T_z-T_x)+c_{bg}\\
\frac{dT_x}{dt}=\frac{n}{4}(K_0
T_z)T_x+\frac{\Gamma_{el}}{3}(T_z-T_x)+c_{bg}
\end{eqnarray}
Here, we have allowed for a difference in the average energy per
particle in the two trap directions, ``$T_z$" and ``$T_x$", so that
$\beta=K_0 T_z + 2 K_1 T_x$.  For the fits, we fix the $d$-dependent
coefficients $K_0$ and $K_1$ using the previous fit to the inelastic
loss rate data in Fig. 1. In addition to heating due to inelastic
loss, we include a measured background heating rate of $c_{bg}=0.01$
$\mu$K/s.  The elastic collision rate in Eqns. 4 and 5 is given by
$\Gamma_{el}=\frac{n \sigma_{el} v}{N_{coll}}$, where the elastic
collision cross section $\sigma_{el}$ is a fit parameter,
$v=\sqrt{\frac{8 k_B (T_z+2 T_x)}{3 \pi \mu}}$, and the constant
$N_{coll}$ can be thought of as the mean number of collisions per
particle required for rethermalization.   We use $N_{coll}=4.1$,
which was computed for $p$-wave collisions~\cite{pwaveretherm},
however, we note that $N_{coll}$ depends on the angular dependence
of the scattering and may be somewhat different for dipolar elastic
collisions.

Supplementary Information is linked to the online version of the
paper at www.nature.com/nature.

\begin{thebibliography}{22}

\bibitem{Krems}
Krems, R. V. Perspective: Cold Controlled Chemistry. \textit{Phys.
Chem. Chem. Phys.} \textbf{10}, 4079 (2008).

\bibitem{DeMille2002}
DeMille, D.  Quantum Computation with Trapped Polar Molecules.
\textit{Phys. Rev. Lett.} \textbf{88,} 067901 (2002).

\bibitem{DeMille}
Andre, A., DeMille, D., Doyle, J. M., Lukin, M. D., Maxwell, S. E.,
Rabl, P., Schoelkopf, R. J., Zoller, P. A coherent all-electrical
interface between polar molecules and mesoscopic superconducting
resonators. \textit{Nature Physics} \textbf{2,} 636-642 (2006).

\bibitem{Yelin} Yelin, S. F., Kirby, K., C\^ote, R. Schemes for robust quantum computation with polar molecules.
\textit{Phys. Rev. A} \textbf{74,} 050301 (2006).

\bibitem{Micheli} Micheli, A., Brennen, G. K., Zoller, P. A toolbox for lattice-spin models with polar molecules. \textit{Nat. Phys.} \textbf{2}, 341-347 (2006).

\bibitem{Lahaye} Lahaye, T., Menotti, C., Santos, L., Lewenstein, M., Pfau, T. The physics of dipolar bosonic quantum gases
\textit{Rep. Prog Phys.} \textbf{72,} 126401 (2009).

\bibitem{Pupillo}
Pupillo, G., Micheli, A., B\"uchler, H. P., Zoller, P. Condensed
Matter Physics with Cold Polar Molecules. in {\it Cold Molecules:
Theory, Experiment, Applications}, ed. Krems, R. V., Stwalley, W.
C., Friedrich, B. (CRC Press, Boca Raton, FL, 2009).

\bibitem{Baranov}
Baranov, M. Theoretical progress in many-body physics with ultracold
dipolar gases. \textit{Phys. Rep.} \textbf{464}, 71-111 (2008).

\bibitem{reviewpolar}
Carr, L. D., DeMille, D., Krems, R. V., Ye. J. Cold and ultracold
molecules: science, technology and applications. \textit{New J.
Phys.} \textbf{11}, 055049 (2009).

\bibitem{polarmol}
Ni, K.-K. \textit{et al.} A High-Phase-Space-Density Gas of Polar
Molecules. \textit{Science} \textbf{322}, 231 (2008).

\bibitem{Micheli2D} Micheli, A. \textit{et al.} Cold polar molecules in two-dimensional traps: Tailoring interactions with external fields for novel quantum phases. \textit{Phys. Rev. A} \textbf{76,} 043604 (2007).

\bibitem{Buchler} B\"uchler, H. P.  \textit{et al.} Strongly correlated 2D quantum phases with cold polar molecules: Controlling the shape of the interaction potential. \textit{Phys. Rev. Lett.} \textbf{98,} 060404 (2007).

\bibitem{Li} Li, Z., Krems, R. V.  Inelastic collisions in an ultracold
quasi-two-dimensional gas. \textit{Phys. Rev. A} \textbf{79}, 050701
(2009).

\bibitem{Quemener}
Qu\'em\'ener, G., Bohn, J. L. Strong Dependence of Ultracold
Chemical Rates on Electric Dipole Moments. \textit{Phys. Rev. A} in
press (2009).

\bibitem{Pfau}
Stuhler, J. \textit{et al.} Observation of Dipole-Dipole Interaction
in a Degenerate Quantum Gas. \textit{Phys. Rev. Lett.} \textbf{95},
150406 (2005).

\bibitem{Stamper-Kurn}
Vengalattore, M., Leslie, S. R., Guzman, J., Stamper-Kurn, D. M.
\textit{et al.} Spontaneously modulated spin textures in a dipolar
spinor Bose-Einstein condensate. \textit{Phys. Rev. Lett.}
\textbf{100}, 170403 (2008).

\bibitem{Goral} Góral, K., Santos, L., Lewenstein, M. Quantum Phases of Dipolar
Bosons in Optical Lattices. \textit{Phys. Rev. Lett}. \textbf{88},
170406 (2002).

\bibitem{Capo} Capogrosso-Sansone, B., Trefzger, C.,
Lewenstein, M., Zoller, P., Pupillo, G. Quantum Phases of Cold Polar
Molecules in 2D Optical Lattices (2009) Preprint at
http://arxiv.org/abs/0906.2009.

\bibitem{pairing} Cooper, N. R., Shlyapnikov, G. V. Stable Topological Superfluid Phase of Ultracold Polar Fermionic Molecules
\textit{Phys. Rev. Lett.} \textbf{103}, 155302 (2009).

\bibitem{hyperfine}
Ospelkaus, S. \textit{et al.} Controlling the Hyperfine State of
Rovibronic Ground State Polar Molecules. \textit{Phys. Rev. Lett.}
in press (2009). Preprint at http://arxiv.org/abs/0908.3931.

\bibitem{ultracoldchem1}
Ospelkaus, S. \textit{et al}. Quantum-State Controlled Chemical
Reactions of Ultracold KRb Molecules. \textit{Science}, in press
(2009).

\bibitem{Kotochigova}
Kotochigova, S. private communication (2009).

\bibitem{Roberts}
Roberts, J. L. Bose-Einstein Condensates with Tunable Atom-atom
Interactions. PhD diss., University of Colorado (2001).

\bibitem{Weber}
Weber, T., Herbig, J., Mark, M., N\"agerl, H.-C., Grimm, R.
Three-Body Recombination at Large Scattering Lengths in an Ultracold
Atomic Gas. \textit{Phys. Rev. Lett.} \textbf{91}, 123201 (2003).

\bibitem{retherm}
Monroe, C. R., Cornell, E. A., Sackett, C. A., Myatt, C. J., Wieman,
C. E. Measurement of Cs-Cs elastic scattering at T=30 $\mu$K.
\textit{Phys. Rev. Lett.} \textbf{70}, 414 (1993).

\bibitem{Bohn}
Bohn, J. L., Cavagnero, M., Ticknor, C. Quasi-universal dipolar
scattering in cold and ultracold gases. \textit{New J. Phys.}
\textbf{11}, 055039 (2009).

\bibitem{Gorshkov} Gorshkov, A. V. \textit{et al}. Suppression of inelastic collisions between polar molecules with a repulsive shield. \textit{Phys. Rev. Lett.} \textbf{101} 073201 (2008).

\bibitem{Demler} Wang, D. W., Lukin, M. D., Demler, E. Quantum fluids of self-assembled chains of polar molecules. \textit{Phys. Rev. Lett.} \textbf{97,} 180413 (2006).

\bibitem{Santos} Klawunn, M., Duhme, J., Santos, L. Bose-Fermi mixtures of self-assembled filaments of fermionic polar molecules. arXiv:0907.4612 (2009).

\bibitem{pwaveretherm}
DeMarco, B., Bohn, J. L., Burke, Jr., J. P., Holland, M. H., Jin, D.
S. Measurement of $p$-wave threshold law using evaporatively cooled
fermionic atoms. \textit{Phys. Rev. Lett.} \textbf{82}, 4208 (1999).

\end{thebibliography}
\end{document}


\section*{Supporting online material}

\section*{Heating due to two-body inelastic $p$-wave collisions}

Loss of trapped atoms or molecules due to inelastic collisions gives
rise to heating through a mechanism called ``anti-evaporation'',
where particles removed from the trap have, on average, a lower
energy than the remaining particles.  Here, we model the heating due
to inelastic two-body $p$-wave collisional loss in a trapped gas, as
relevant to our experiments.

To accommodate the spatial
anisotropy of the dipole-dipole interaction and the possibility of
an anisotropic energy distribution in the trapped gas, we define
``temperatures'' along three spatial directions using
\begin{eqnarray}
T_i=\frac{E_i}{k_B N},
\end{eqnarray}
where the index $i=x,y,z$ and $N$ is the number of molecules.  As
the trap shape is a flat disk perpendicular to the vertical ($z$)
axis, which is also the direction of the electric field, it is
natural to assume that the temperatures in the two radial trap
directions, $T_x$ and $T_y$, are equal. The energies $E_i$ are
defined as
\begin{eqnarray}
E_z=\sum\left(\frac{1}{2}m v_z^2+\frac{1}{2}m\omega_z z^2\right)\\
E_x=\sum\left(\frac{1}{2}m v_x^2+\frac{1}{2}m\omega_x
x^2\right)\nonumber,
\end{eqnarray}
where $m$ is the particle mass, $\omega_z$ and $\omega_r$ are the
axial and radial trapping frequencies, respectively, and the sum is
taken over all particles in the trap. The heating rates are then
given by
\begin{eqnarray}
\frac{dT_i}{dt}    = -\frac{T_i}{N}\frac{dN}{dt}+\frac{1}{k_B
N}\frac{dE_i}{dt}.
\end{eqnarray}
Each two-body inelastic collision removes two particles, so the loss
rate is given by
\begin{eqnarray}
\frac{dN}{dt}= -2 N \Gamma_{coll},
\end{eqnarray}
where $\Gamma_{coll}$ is the inelastic collision rate per particle.
Similarly,
\begin{eqnarray}
\frac{dE_i}{dt}= -\Delta E_i N \Gamma_{coll},
\end{eqnarray}
where $\Delta E_i$ is the average $E_i$ for a pair of particles that
is lost from the trap. Putting this together, we have
\begin{eqnarray}
\frac{dT_i}{dt}    = 2 T_i \Gamma_{coll}-\frac{\Delta E_i}{k_B}
\Gamma_{coll}.
\end{eqnarray}

To find the heating caused by inelastic collisional loss we need to
calculate $\Delta E_z$ and $\Delta E_x$. The total energy of any
pair of particles is the sum of four terms: kinetic and potential
energy of the center-of-mass motion of the two colliding particles,
and their relative kinetic and potential energy.  In any of the
three directions, the ensemble average for the energy is $2k_BT$
with equal contributions ($1/2 k_BT$) coming from each of the four
terms. However, the average energy for pairs of particles that
undergo inelastic collisions is, in general, different from the
ensemble average because the collision rate depends on the relative
position and relative velocity. Because collisions are only
sensitive to relative motion, the kinetic and potential energy of
the center-of-mass motion of the two colliding particles will be on
average the same as the ensemble average, ($1/2 k_BT$). The relative
potential energy is zero for colliding particles (since they must be
at the same position to collide). This is the reason that
collisional loss in a trap leads to``anti-evaporative'' heating, in
contrast to the elastic-collision-based evaporative cooling where
energetic particles are preferentially removed from a trap.

It remains then to consider the relative kinetic energy.  For
$p$-wave inelastic collisions, the collision rate increases linearly
with the relative energy of the colliding particles.  Furthermore,
as discussed in the main text, the spatial anisotropy of the
dipole-dipole interaction can result in different rates for $m_L=0$
and $m_L=\pm1$ collisions.  Therefore, we calculate the heating for
$m_L=0$ and $m_L=\pm1$ collisions separately and then add the
heating rates in the final result. For $m_L=0$ collisions, we find
the average relative energy for two colliding particles using
\begin{eqnarray}
E_{z,rel}^0 &=& \frac{1}{2}\mu\frac{\int_{-\infty}^\infty \!
 f(v_z,v_r) v_z^4 dv_z}{\int_{-\infty}^\infty \!
 f(v_z,v_r) v_z^2 dv_z}\nonumber\\
 &=& \frac{3}{2} k_B T_z\\
E_{x,rel}^0&=&\frac{1}{4}\mu\frac{\int_{0}^\infty \!
 f(v_z,v_r) v_r^2 (2 \pi v_r) dv_r}{\int_{0}^\infty \!
 f(v_z,v_r) (2 \pi v_r) dv_r}\nonumber\\
 &=& \frac{1}{2} k_B T_x
\end{eqnarray}
where we assume a Maxwell-Boltzmann distribution of relative
velocities given by
\begin{eqnarray}
f(v_z,v_r)\propto \exp \left(-\frac{\mu v_r^2}{2 k_B T_r}\right)
\exp \left(-\frac{\mu v_z^2}{2 k_B T_z}\right)
\end{eqnarray}
with $v_r^2=v_x^2+v_y^2$. A factor of $v_z^2$ in both the numerator
and denominator of Eqn. (7) accounts for the fact that the $m_L=0$
$p$-wave inelastic collision rate scales as collision energy
$\propto v_z^2$.  Similarly, for $m_L=\pm1$ collisions, the
collision rate scales as $v_r^2$ and we find the average relative
energy for two colliding particles using
\begin{eqnarray}
E_{z,rel}^1&=&\frac{1}{2}\mu\frac{\int_{-\infty}^\infty \!
 f(v_z,v_r) v_z^2 dv_z}{\int_{-\infty}^\infty \!
 f(v_z,v_r) dv_z}\nonumber\\
 &=& \frac{1}{2} k_B T_z\\
 E_{x,rel}^1 &=& \frac{1}{4}\mu\frac{\int_{0}^\infty \!
 f(v_z,v_r) v_r^4 (2 \pi v_r) dv_r}{\int_{0}^\infty \!
 f(v_z,v_r) v_r^2 (2 \pi v_r) dv_r}\nonumber\\
 &=&  k_B T_x.
\end{eqnarray}
Adding this to the $k_B T_i$ from the center-of-mass energies, we
get
\begin{eqnarray}
\Delta E_{z}^0&=& \frac{5}{2} k_B T_z\\
\Delta E_{x}^0&=& \frac{3}{2} k_B T_x\\
\Delta E_{z}^1&=& \frac{3}{2} k_B T_z\\
\Delta E_{x}^1&=& 2 k_B T_x.
\end{eqnarray}

Putting this into the expression for the heating rate (Eqn. (6)), we
find that
\begin{eqnarray}
\frac{dT_z}{dt}&=& -\frac{1}{2} T_z \Gamma_{coll}^0+\frac{1}{2} T_z \Gamma_{coll}^1\\
\frac{dT_x}{dt}&=& +\frac{1}{2} T_x \Gamma_{coll}^0.
\end{eqnarray}

In the main text, we defined the loss rate coefficient as
\begin{eqnarray}
\beta&=&K_0 T_z+2 K_1 T_x\\
 &=&\frac{2}{n}\left(\Gamma_{coll}^0+\Gamma_{coll}^1\right)\nonumber,
\end{eqnarray}
where $K_0$ and $K_1$ depend on the induced dipole moment. The
inelastic collision rate per particle is related to $\beta$ through
\begin{eqnarray}
\Gamma_{coll}^0=\frac{K_0 T_z n}{2}\\
\Gamma_{coll}^1=\frac{2 K_1 T_x n}{2}.
\end{eqnarray}
With this substitution, our final result for the heating due to
inelastic loss is
\begin{eqnarray}
\frac{dT_z}{dt}&=& \frac{n}{4} (-K_0 T_z+ 2 K_1 T_x) T_z\\
\frac{dT_x}{dt}&=& \frac{n}{4} (K_0 T_z) T_x.
\end{eqnarray}

If we assume that the ensemble stays in cross-dimensional
equilibrium, $T=T_z=T_x$, then the overall heating rate is simply
\begin{eqnarray}
\frac{dT}{dt}&=& \frac{1}{3}\frac{dT_z}{dt}+\frac{2}{3}\frac{dT_x}{dt}\nonumber\\
  &=& \frac{n}{12} (K_0+2K_1) T^2,
\end{eqnarray}
or equivalently,
\begin{eqnarray}
\frac{dT}{dt}/(T^2 n)=\frac{\beta/T}{12}.
\end{eqnarray}